\documentclass[prd,preprint,showpacs,groupedaddress]{revtex4-1}
\usepackage{amsmath}
\usepackage{amsfonts}
\usepackage{amssymb}
\usepackage{geometry}
\usepackage{natbib}
\usepackage[english]{babel}
\usepackage{graphicx}
\usepackage{epstopdf}
\usepackage{caption}
\begin{document}
\captionsetup[figure]{labelfont={bf},labelformat={default},labelsep=period,name={Fig.}}

  \title{Thermodynamics of the Bardeen Black Hole in Anti-de Sitter Space}

  \author{Cong Li} \author{Chao Fang} \author{Miao He} \author{Jiacheng Ding} \author{Jianbo Deng}\email[Corresponding author:]{dengjb@lzu.edu.cn}

  \affiliation{Institute of Theoretical Physics, LanZhou University,
    Lanzhou 730000, P. R. China}

  \date{\today}

  \begin{abstract}

  	 In this article we study thermodynamics of the regular black holes with Bardeen-AdS black hole. The cut-off radius which is the minimal radius of the stable Bardeen-AdS black hole have been got from temperature and heat capacity analysis respectively. Moreover, the thermodynamical stability of the Bardeen-AdS black hole is learned by the Gibbs free energy and the heat capacity. In this work we find the similar characteristics to the Van der Waals liquid-gas system.
  \end{abstract}

  %\pacs{04.20.Fy, 04.50.Kd, 04.60.Ds}


  \maketitle

  \section{\label{sec:level1}INTRODUCTION}

Lots of interesting phenomena have been presented by physicists after many years of researches about black hole thermodynamics. Since Hawking and Page\cite{1} firstly studied the thermodynamic properties of AdS black holes and found the certain phase transition in the phase space, the thermodynamic properties of different black holes, such as Kerr-Newman black holes in AdS space\cite{2}, regular black holes\cite{3} and charged AdS black holes\cite{4} have been discussed. Moreover, it was shown that such phase transitions appeared not only in AdS space, but also the asymptotically dS space and flat space\cite{5}. On the other hand, thermodynamics of black holes in AdS space has drawn more attention due to the AdS/CFT (conformal field theory) duality\cite{6}. AdS/CFT duality led to a new way for studying the phase transitions of AdS black holes in the dual field\cite{7}.
\par
Generally, black holes have singularities inside the horizons where the curvature scalar goes to infinity, which has inspired many researchers to construct the singularity-free black holes, called regular black holes. Bardeen in 1968 obtained a black hole solution without a singularity which was now well known as the Bardeen black hole\cite{8}. In 2000, Ay\'{o}n-Beato and Garc\'{i}a showed that the Bardeen black hole could be interpreted as a gravitationally collapsed magnetic monopole arising in a specific form of nonlinear electrodynamics\cite{9} and gave the corresponding Lagrangian for the suggested nonlinear electrodynamics. Therefore, the stress energy tensor of nonlinear electrodynamics worked as the source of Einstein field equations. In recent years, Bardeen black hole has drawn more attention. Moreno and Sarbach\cite{10} studied the dynamical stability properties of the Bardeen black hole and other regular black holes. The thermodynamical stability of black holes with nonlinear electromagnetic fields was researched in~\cite{Breton:2014nba}. Quasinormal modes of the Bardeen black hole have also been studied by~\cite{11}\cite{12}. And the thermodynamic quantities of the Bardeen black hole have been addressed in~\cite{13}\cite{Saleh:2017vui}\cite{Sharif:2011ja}. Moreover, Sharmanthie Fernando studied the Bardeen black hole in de-Sitter space\cite{fernando2017bardeen}.\par

\section{the Bardeen black holes in AdS space}
The Bardeen black hole is known as a black hole without singularity, and it can  be interpreted as a gravitationally collapsed magnetic monopole arising in a specific form of non-linear electrodynamics~\cite{9}.The corresponding action with $\Lambda$ term can be given by
  \begin{equation}
    S=\int d^4x\sqrt{-g'}\left[\frac{(R-2\Lambda)}{16\pi}-\frac{1}{4\pi}\mathcal{L}(F)\right].
  \end{equation}
 Here, $R$ is the scalar curvature, $g'$ is the determinant of the metric tensor, $\Lambda$ is the cosmological constant and $\mathcal{L}(F)$~\cite{9}is a function of $F=\frac{1}{4}F_{\mu\nu}F^{\mu\nu}$ given by
  \begin{equation}
    \mathcal{L}(F)=\frac{3}{2sg^2}\left(\frac{\sqrt{2g^2F}}{1+\sqrt{2g^2F}}\right)^{\frac{5}{2}}.
  \end{equation}
  The parameter $s$ in the above equation is given by $\frac{|g|}{2M}$ where $g$ and M corresponds to the magnetic charge and the mass of the black hole. The field strength of the non-linear electrodynamics is $F_{\mu\nu}=2\nabla_{[\mu}A_{\nu]}$.
  The field equations of motion derived from the action in Eq.(1) is given by
  \begin{equation}
    G_{\mu\nu}+\Lambda g'_{\mu\nu}=2\left(\frac{\partial\mathcal{L}(F)}{\partial F}F_{\mu\lambda}F^{\lambda}_{\nu}-g'_{\mu\nu}\mathcal{L}(F)\right),
  \end{equation}
  \begin{equation}
    \nabla_{\mu}\left(\frac{\partial\mathcal{L}(F)}{\partial F}F^{\nu\mu}\right)=0,
  \end{equation}
  \begin{equation}
    \nabla_{\mu}(*F^{\nu\mu})=0.
  \end{equation}
 Then, we consider the line element of the Bardeen-AdS black hole as
  \begin{equation}
    \label{eq:element}
    ds^2=-f(r)dt^2+\frac{1}{f(r)}dr^2+r^2d\Omega^2,
  \end{equation}
  where $f(r)=1-\frac{2m(r)}{r}$ and $d\Omega^2$ stands for the standard element on the unit two-sphere.
  Following the ansatz for Maxwell equation $F_{\mu\nu}=2\delta^{\theta}_{[\mu}\delta^{\phi}_{\nu]}Z(r,\theta)$~\cite{9}and with the help of Eq.~(4), we can get
  \begin{equation}
    \label{F}
    F_{\mu\nu}=2\delta^{\theta}_{[\mu}\delta^{\phi}_{\nu]}g(r)sin\theta.
  \end{equation}
  Using the condition $dF=g(r)sin\theta~dr\wedge d\theta \wedge d\phi=0$, we conclude that $g(r)=const.=g$. Hence, the field strength is $F_{\theta\phi}=-F_{\phi\theta}=gsin\theta$ with $F=\frac{g^2}{2r^4}$. Substituting the expression in Eq.~(2), we get
  \begin{equation}
    \label{L}
    \mathcal{L}(F)=\frac{3Mg^2}{(g^2+r^2)^{\frac{5}{2}}}.
  \end{equation}
  Moreover, the $^t_{t}$ component of Eq.~(3) can yield a solution for $m(r)$ with
  \begin{equation}
    \label{m}
    m(r)=\frac{Mr^3}{(g^2+r^2)^\frac{3}{2}}-\frac{r^3}{2l^2}.
  \end{equation}
  So the metric of Bardeen-AdS black hole is
  \begin{equation}
    \label{f(r)}
    f(r)=1-\frac{2Mr^2}{(g^2+r^2)^\frac{3}{2}}+\frac{r^2}{l^2}.
  \end{equation}
  It would be convenient to write the metric $f(r)$ as
  \begin{equation}
    \label{f(x)}
    f(a)=1-\frac{2ma^2}{(Q^2+a^2)^\frac{3}{2}}+a^2,
  \end{equation}
  where $a=\frac{r}{l},~~m=\frac{M}{l},~~Q=\frac{g}{l}$. Then, we can get the minimum radius with $f(a)=\frac{\partial f(a)}{\partial a}=0$ and $\frac{\partial f(a)^2}{\partial^2 a}|_{a_{c}}>0$, which is
  \begin{equation}
    \label{a_{0}}
    a_{c}=\frac{\sqrt{-1+\sqrt{1+24Q^2}}}{\sqrt6}.
  \end{equation}
  When
  \begin{equation}
     f(r_{h})=0,
  \end{equation}
  we get
  \begin{equation}
    \label{eq:M}
    M=(1+\frac{r^2_{h}}{l^2})\frac{(r^2_{h}+g^2)^\frac{3}{2}}{2r^2_{h}}.
    \end{equation}
  The pressure in AdS black holes corresponds to the $\Lambda$ with
  \begin{equation}
  P=-\frac{\Lambda}{8\pi}=\frac{3}{8\pi l^2},
  \end{equation}
  and the Hawking temperature $T$ can be obtained from the first derivative of $f(r)$ at the horizon~\cite{21}
  \begin{equation}
    \label{eq:T}
    T=\frac{\kappa}{2\pi}=\frac{f'(r_{h})}{4\pi}=\frac{3r_{h}}{4\pi(r_{h}^2+g^2)}+\frac{3r_{h}^3}{4\pi(r_{h}^2+g^2)l^2}-\frac{1}{2\pi r_{h}}.
  \end{equation}
   It can come back to Schwarzschild AdS black holes~\cite{1} when $g=0$. It can also back to Bardeen black hole without cosmological constant~\cite{Sharif:2011ja}. The entropy $S$ of the black hole satisfies
  \begin{equation}
    \label{eq:S}
    S=\frac{A}{4}=\pi r_{h}^2,
  \end{equation}
  where A is the area of the horizon.
  \begin{figure}[htp]
  \centering
  \includegraphics[width=8cm]{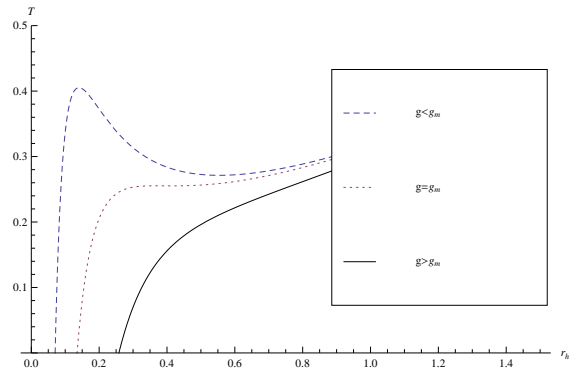}
  \caption{Plot of the Hawking temperature respect to $r_{h}$. The $g$ increase from top to bottom. We have set $l=1$, and $g=0.05, 0.0986571, 0.2$ from top to bottom.}
  \end{figure}
\subsection{Hawking temperature}
As can be seen from FIG.1, when $g$ is large enough, the Hawking temperature respect to $r_{h}$ is increasing monotonically. It is interesting that rich phenomena will be showed when $g$ becomes small enough. There is a critical value of $g$. Using
\begin{equation}
\frac{\partial T}{\partial r_{h}}=0,\frac{\partial^2T}{\partial^2r_{h}}=0,
\end{equation}
and a appropriate solution is
\begin{equation}
g_{m}=\frac{\sqrt{219-13\sqrt{273}}}{12\sqrt{3}}l.
\end{equation}
The others are negative or not real. $g_{m}$ is a critical value, that is to say, when $g>g_{m}$, there is just a black hole without the second-order phase transition. Phase transition will be turned on while $g\le g_{m}$.
\par So the Bardeen-AdS black hole can be strongly influenced by parameter $g$. There is also another phenomenon that the Bardeen-AdS black hole can not be infinitesimal, one can also easily calculate the minimal radius, when
\begin{equation}
T=0,r_{s}=\frac{\sqrt{\sqrt{l^2(24g^2+l^2)}-l^2}}{\sqrt{6}},
\end{equation}
 which is same with Eq.~(12). In our article, we call $r_{s}$ cut-off radius. When one get a radius $r_{h}$ of Bardeen-AdS black holes by other method, the rationality of it should be discussed in our framework. A comparison with $r_{m}$ is necessary. If it is smaller than the cut-off radius, the discussion of the black hole of this state would be meaningless.
\subsection{\label{app:subsec}Gibbs free energy}
In this subsection, we study the Gibbs free energy in grand-canonical ensemble. Because the black hole mass M can be regarded as the enthalpy H in AdS space\cite{22}, the Gibbs free energy can be written as
\begin{equation}
G=M-TS.
\end{equation}
Then we calculate out:
\begin{equation}
G=\frac{(g^2+r_{h}^2)^\frac{3}{2}(1+\frac{r_{h}^2}{l^2})}{2r_{h}^2}-\frac{r_{h}(-2g^2l^2+l^2r_{h}^2+3r_{h}^4)}{4l^2(g^2+r_{h}^2)}.
\end{equation}
A figure of $G$ respect to $r_{h}$ is plotted as Figure 2.
\begin{figure}[htp]
\centering
\includegraphics[width=8cm]{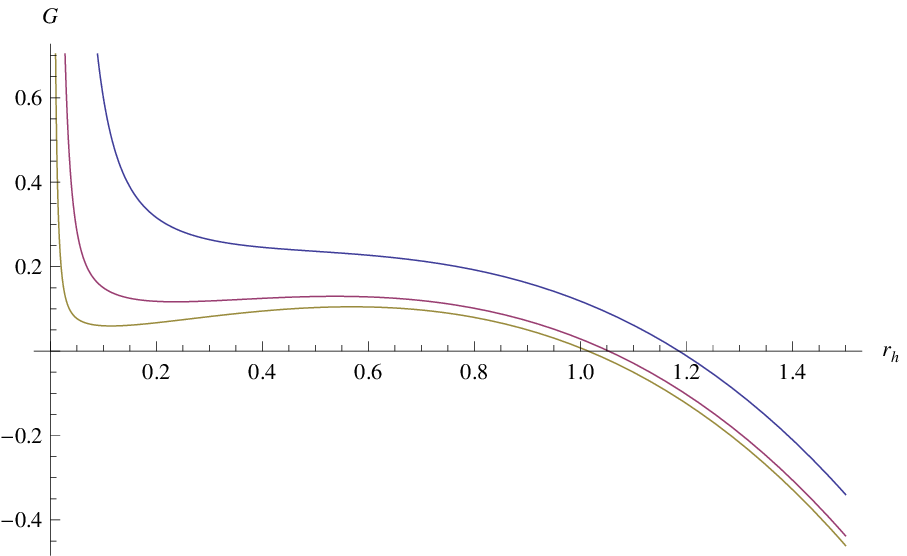}
\caption{Plot of the Gibbs free energy of $r_{h}$. The $g$ decrease from top to bottom. We have set $l=1$, and $g=0.2, 0.0986571, 0.05$ from top to bottom.}
\end{figure}
As can be seen from FIG. 2, there is no extreme value point for $g>g_{m}$. As for $g<g_{m}$, there are two extreme value points. And there is only an inflection point for $g=g_{m}$. It is obvious that the value of $g$ has something to do with the phase transitions. In all the cases, the curves cross the horizontal axis and $G$ change from positive to negative. And second-order phase transition will occur only for $g\le g_{m}$. We will study these carefully in next subsection. The Gibbs free energy becomes very large when $r_{h}$ is small. It is well known that when the Gibbs free energy is lower, the system is more stable and vice versa. So, the Bardeen-AdS black hole is very unstable when radius of event horizon is small. It is consistent with the result we have illustrated before. Conversely, when the Bardeen-AdS black hole is large enough, it will be stable.
\subsection{\label{app:subsec}The heat capacity}
The thermodynamical stability is usually studied by analysing the behaviour of the entropy $S$. The stability requires that the entropy hypersurface lies everywhere below its family of tangent hyperplanes, i.e. the entropy must be a concave function of the entropic extensive parameters.  This requires thermal capacity must be positive. So, in this section, we study the heat capacity.
There are three cases for $g<g_{m}$, $g=g_{m}$ and $g>g_{m}$. We will discuss respectively in this subsection. Generally, the heat capacity is obtained as
\begin{equation}
C_{p}=T\frac{\partial S}{\partial T}|_{p}=\frac{2\pi r_{h}^2(g^2+r_{h}^2)(-2g^2l^2+l^2r_{h}^2+3r_{h}^4)}{2g^4l^2-l^2r_{h}^4+3r_{h}^6+g^2(7l^2r_{h}^2+9r_{h}^4)}.
\end{equation}
When $C_{p}=0$, there are four solutions, the reasonable one that we get is
\begin{equation}
r_{s}=\frac{\sqrt{\sqrt{l^2(24g^2+l^2)}-l^2}}{\sqrt{6}},
\end{equation}
which is the same with Eq.~(12) and Eq.~(20). As can be seen before, $r_{s}$ is cut-off radius, and the Bardeen-AdS black hole is unstable when $r_{h}<r_{s}$. From the heat capacity, we all know that black holes are stable when $C_{p}>0$. which is the classical thermodynamical stability condition\cite{Peca:1998cs}. So we get the same conclusion through both Hawking temperature and heat capacity. \par
When the heat capacity is divergent, the second-order phase transition will occur. We can calculate it by setting
\begin{equation}
2g^4l^2-l^2r^4+3r^6+g^2(7l^2r^2+9r^4)=0.
\end{equation}
When $0<g<\sqrt{\frac{73l^2}{144} - \frac{13}{144}\sqrt{\frac{91}{3}}l^2}$ and $l > 0$, the function has two real positive solutions, which are
\begin{equation}
\begin{split}
r_{1}=(&-g^2+\frac{l^2}{9}+\frac{9\times2^\frac{1}{3}g^4}{A}-\frac{9\times2^\frac{1}{3}g^2l^2}{A}\\&+\frac{2^\frac{1}{3}l^4}{9A}+\frac{1}{9\times2^\frac{1}{3}A})^\frac{1}{2},
\end{split}
\end{equation}
\begin{equation}
\begin{split}
r_{2}=(&-g^2+\frac{l^2}{9}-\frac{9g^4}{2^\frac{2}{3}A}+\frac{9i\sqrt{3}g^4}{2^\frac{2}{3}A}+\frac{9g^2l^2}{2^\frac{2}{3}A}\\&-\frac{9i\sqrt{3}g^2l^2}{2^\frac{2}{3}A}-\frac{l^4}{9\times2^\frac{2}{3}A}+\frac{il^4}{3\times2^\frac{2}{3}\sqrt{3}A}\\&-\frac{1}{18\times2^\frac{1}{3}A}-\frac{1}{6\times2^\frac{1}{3}\sqrt{3}iA})^\frac{1}{2},
\end{split}
\end{equation}
where
\begin{equation}
\begin{split}
A=(&-1458g^6+1701g^4l^2-243g^2l^4+2l^6\\&+27(1944g^{10}l^2-3915g^8l^4+1990g^6l^6\\&-19g^4l^8)^\frac{1}{2})^\frac{1}{3}.
\end{split}
\end{equation}
It is obvious that there are two divergent points. And we plot pictures to study in detail.
\begin{figure}[htp]
\centering
\includegraphics[width=8cm]{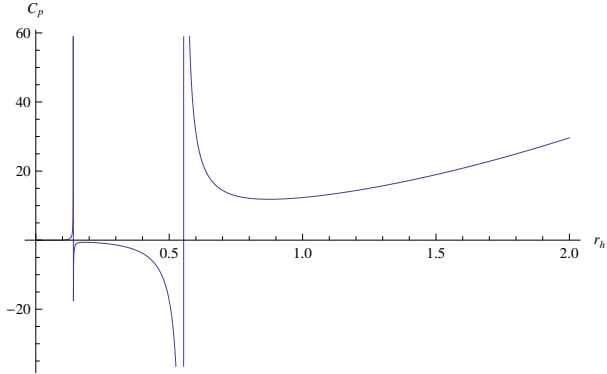}
\caption{Plot of the heat capacity respect to $r_{h}$. We have set $l=1$, and $g=0.05$.}
\end{figure}
\begin{figure}[htp]
\centering
\includegraphics[width=8cm]{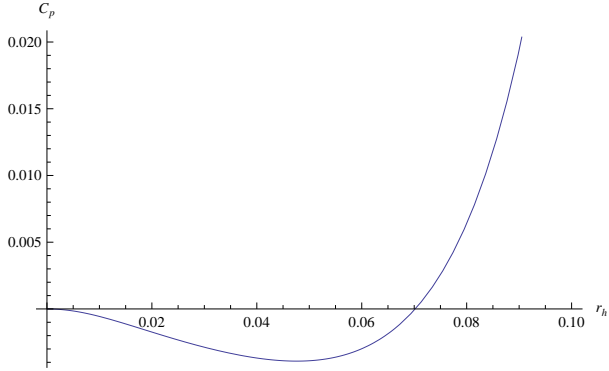}
\caption{Partial enlarged drawing of FIG. 3 at first divergent point.}
\end{figure}
\par As can be seen from FIG. 4, when $r_{h}$ is smaller than $r_{m}$ the heat capacity is negative, which also shows the minimal radius is $r_{m}$. The two divergent points are also shown in FIG. 3. This means the second-order phase transitions will occur at those points. The heat capacity changes from  positive to negative at the first divergent point, the case at the second point is inverse. Generally, the volume of black holes is not changed while second-order phase occurs. It is just from stable state to unstable state and vice versa. Between the two divergent points, the heat capacity is negative, this is an intermediate state which is unstable. The detailed information about this state needs to be researched in the future, and ref.~\cite
{16} can give some advice.
\par When $r_{1}=r_{2}$, one can solve
\begin{equation}
g=\frac{\sqrt{219-13\sqrt{273}}}{12\sqrt{3}}l.
\end{equation}
In this case, two phase transition points degenerate into one. That means it is the critical phase transition. So, it is not strange that Eq.~(19) and Eq.~(29) have the same result.
\begin{figure}[htp]
\centering

\includegraphics[width=8cm]{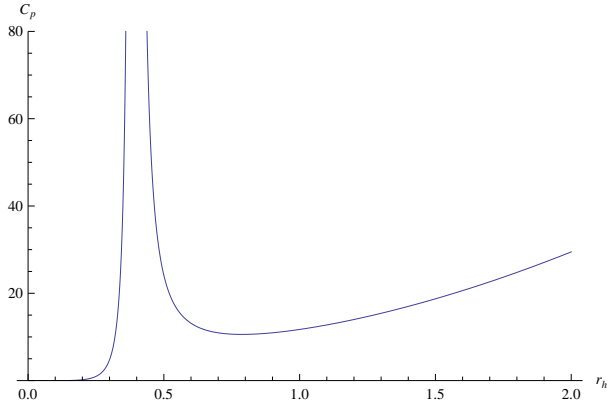}
\caption{Plot of the heat capacity of $r_{h}$. We have set $l=1$, and $g=0.0986571$.}
\end{figure}
\begin{figure}[htp]
\centering
\includegraphics[width=8cm]{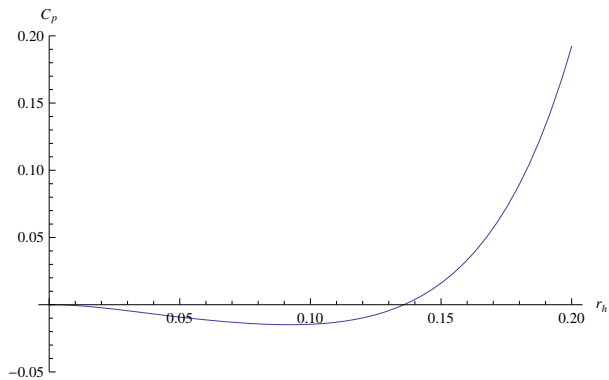}
\caption{Partial enlarged drawing of FIG. 5 at divergent point.}
\end{figure}
\begin{figure}[htp]
\centering
\includegraphics[width=8cm]{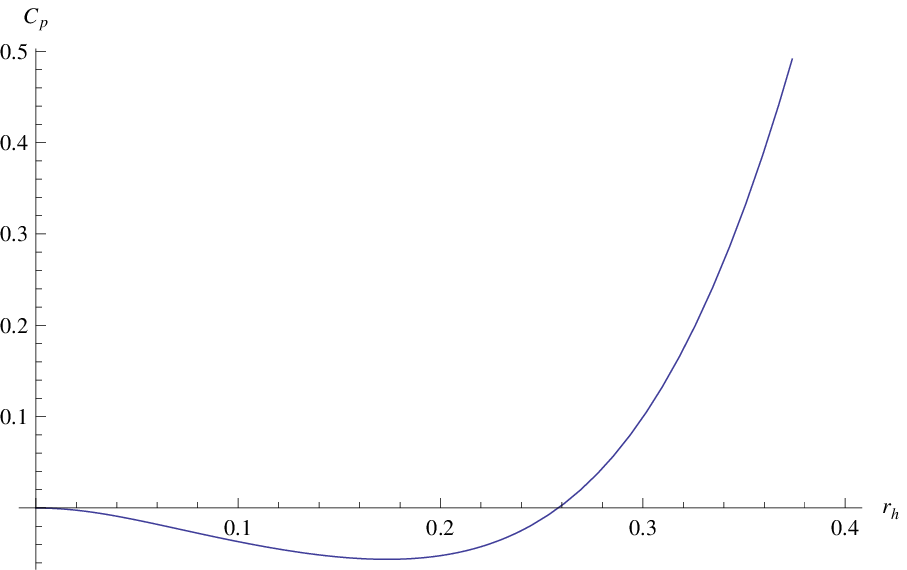}
\caption{Plot of the heat capacity of $r_{h}$. We have set $l=1$, and $g=0.2$.}
\end{figure}
\par As can be seen from FIG. 5, there is only one divergent point. We can also see from FIG. 6, when $r_{h}$ is smaller than $r_{m}$ the heat capacity is negative, and positive for $r_{h}>r_{m}$. After phase transition, it is also positive. The physics process presents the change from the small black hole to the large black hole~\cite{17}.
\par The last case is that when $g>g_{m}$, there is no second-order phase transition.

As can be seen from FIG. 7, there is no divergent point. And when $r>r_{s}$, the heat capacity is positive, so the Bardeen black hole is stable.
\section{\label{sec:level1}conclusions}
Phase transitions for AdS black holes have been analyzed in thermodynamic variables. We study the Bardeen-AdS black hole, which has no singularity with a parameter $g$. We find the cut-off radius $r_{s}$. When $r_{h}$ is small, the Bardeen-AdS black hole is very unstable. This can be also read from Gibbs free energy and the heat capacity. Moreover, different phenomena will turn on when the value of $g$ is different. We computed the critical value $g_{m}$. When $g<g_{m}$, there are two second-order phase transition points. Between the two points, the capacity is negative, which is the unstable intermediate state. When $g=g_{m}$, the two phase transition points degenerate into one, which represents the change from a small black hole to a large black hole. There is no second-order phase transition for $g>g_{m}$. The stability of the Bardeen-AdS black hole is also discussed from Gibbs free energy and the heat capacity, both of them point to the same result.

  \section*{Conflicts of Interest}
  The authors declare that there are no conflicts of interest regarding the publication of this paper.

  \section*{Acknowledgments}
  We would like to thank the National Natural Science Foundation of China~(Grant No.11571342) for supporting us on this work.

  \section*{References}
 \bibliographystyle{unsrt}
 \bibliography{article}
\end{document}